\documentclass{aa}  

\usepackage{graphicx}
\usepackage{txfonts}
\usepackage[colorlinks=true, linkcolor=blue, citecolor=blue, urlcolor=blue]{hyperref}
\usepackage{indentfirst}  
\usepackage{longtable}
\usepackage{appendix}
\usepackage{caption} 
\usepackage{lastpage}
\usepackage{threeparttable} 
\usepackage{comment}
\usepackage{color}
\usepackage[dvipsnames,table]{xcolor}

\newcommand\newtext[1]{{{#1}\sl}}
\newcommand\gaia{\text{\it Gaia}\xspace}

\begin{document} 

\setlength{\parindent}{15pt} 

   \title{An ancient L- type family associated to (460)~Scania in the Middle Main Belt as revealed by \gaia DR3 spectra}

   \author{R. Balossi\inst{1}
          \and 
          P. Tanga\inst{1}
          \and 
          M. Delbo \inst{1, 2}
          \and
          A. Cellino \inst{3}
          \and
          \newtext{F. Spoto} \inst{4}
          }

   \institute{Université Côte d'Azur, Observatoire de la Côte d'Azur, CNRS, Laboratoire Lagrange, Bd de l'Observatoire, CS 34229, 06304 Nice Cedex 4, France\\
   \email{roberto.balossi@oca.eu}
   \and
            University of Leicester, School of Physics and Astronomy, University Road, LE1 7RH, Leicester, UK
   \and
            INAF - Osservatorio Astrofisico di Torino, via Osservatorio 20, 10025, Pino Torinese (TO), Italy
    \and
            \newtext{Minor Planet Center, Smithsonian Astrophysical Observatory, 60 Garden Street, Cambridge, MA 02138, USA}
   }
   \date{}

  \abstract
   {Asteroid families are typically identified using hierarchical clustering methods (HCM) in the proper element phase space. However, these methods struggle with overlapping families, interlopers, and the detection of older structures. Spectroscopic data can help overcome these limitations. The \gaia Data Release 3 (DR3) contains reflectance spectra at visible wavelengths for 60,518 asteroids over the range between 374–1034 nm, representing a large sample that is well suited to studies of asteroid families.} 
   {Using \gaia spectroscopic data, we investigate a region in the central Main Belt centered around 2.72 AU, known for its connection to L- type asteroids. Conflicting family memberships reported by different HCM implementations underscore the need for an independent dynamical analysis of this region.}
   {We determine family memberships by applying a color taxonomy derived from \gaia data and by assessing the spectral similarity between candidate members and the template spectrum of each family.}
   {We identify an L-type asteroid family in the central Main Belt, with (460) Scania as its largest member. Analysis of the family's V-shape indicates that it is relatively old, with an estimated age of approximately 1 Gyr, which likely explains its non-detection by the HCM. The family's existence is supported by statistical validation, and its distribution in proper element space is well reproduced by numerical simulations. Independent evidence from taxonomy, polarimetry, and spin-axis obliquities consistently supports the existence of this L-type family.}
  {This work highlights the value of combining dynamical and physical data to characterize asteroid families and raises questions about the origin of L-type families, potentially linked to primordial objects retaining early protoplanetary disk properties. Further spectroscopic data are needed to clarify these families.}


   \maketitle

\section{Introduction} \label{Introduction}

Although collisions among asteroids are rare, they have occurred frequently over the age of the Solar System, playing a key role in shaping the Main Belt. Energetic collisions form what are known as asteroid families. Families are initially very compact in the space of the osculating elements, but over time they evolve and diffuse under the action of gravitational perturbations, non-gravitational effects and further collisional erosion.

Asteroid families are usually identified in the phase space of proper elements using hierarchical clustering methods (HCM, \citealt{zappala-1990}). Several HCM implementations are found in the literature (\citealt{novakovic-2022} and references therein). These usually differ in the approach implemented to define the cutoff velocity ($V_{cut}$), which is used as an upper limit to the most distant family members. 
If the cutoff velocity is too large, the family members tend to merge with the background population or to link into nearby families, a limitation known as the "chaining effect" (\citealt{novakovic-2022}). In addition, HCM implementations also fail to identify old families since they are too dispersed in the proper elements phase space for these algorithms to work properly. Moreover, HCM can mistakenly associate unrelated objects with a family simply because they happen to have similar proper elements. The fraction of interlopers in a family identified solely through HCM has been estimated in the past to be around 10\% (\citealt{migliorini-1995}).

Family members disperse in space due to non-gravitational forces, such as the Yarkovsky effect (\citealt{bottke-2006}), that change the asteroid's semi-major axis at an average rate $da/dt$ inversely proportional to the diameter $D$. Prograde rotating asteroids have $da/dt>0$, and retrograde ones have $da/dt<0$. \newtext{This creates correlated distributions of family members in the ($a$, $1/H$) and ($a$, $1/D$) planes, called V-shapes. The ''V'' slope ($K$) is used to derive family ages (\citealt{vokrouhlicky-2006}, \citealt{spoto-2015}, \citealt{ferrone-2023}). An additional modeling of the YORP effect would be important for estimating the role of the initial velocity ejection field on top of the family age. In simpler formulations, when only the Yarkovsky effect is included, the estimated age is rather an upper bound. This, however, is not a critical issue for old families like the one analyzed in this work.} Asteroids, as they drift in semi-major axis, encounter orbital resonances with the planets that change their orbital eccentricity $e$ and inclination $i$, but not their semi-major axis $a$. Significant changes in eccentricity may lead to close planetary encounters, potentially ejecting objects from the family over relatively short timescales. Thus, families become harder to identify as their age increases because they are more and more dispersed (\citealt{parker-2008}) and they overlap with each other. It has been shown that the V-shape can be used to identify old families with strongly dispersed ($e$, $i$), thus invisible to the HCM (\citealt{bolin-2017}, \citealt{delbo-2019}, \citealt{delbo-2017}, \citealt{ferrone-2023}, \citealt{walsh-2013}).

In this study, we focus on a low-inclination region in the middle belt centered around 2.72~AU where there is a reported abundance of L- type asteroids and conflicting indications of the presence of L- type families. Thanks to the SMASS spectroscopy \citep{bus&binzel-2002} some L-type asteroids were identified close to (2085)~Henan. By using the HCM method \citet{broz-2013} tentatively reported a cluster of 946 asteroids with (2085)~Henan as its largest member, too dispersed to be confirmed as a family. \newtext{\cite{nesvorny-2024}} identified in this region (using a $V_{cut}$=50 m/s) a large Henan family of 1872 members. At the same time, with a more conservative linking, \citet{milani-2014} only identifies a few minor clusters partially overlapping the same volume of the orbital elements space, none associated with (2085)~Henan. Recent classifications maintained for instance at the Asteroid Family Portal (AFP)\footnote{\url{http://asteroids.matf.bg.ac.rs/fam/}} (\citealt{novakovic-2022}) also do not report the so-called Henan family. 

This intriguing situation is associated with the interest that L-type objects may deserve, as at least some are peculiar in their polarimetric properties. Such asteroids were nicknamed ''Barbarians'' from the first discovered prototype of the category, (234) Barbara \citep{cellino-2006, gilhutton-2008, gilhutton-2014}. 
Their reflectance spectra in the visible and near-IR have suggested a link to a possible large abundance of calcium-aluminum rich inclusions (CAIs), embedded in a CV-like matrix with a low degree of thermal or aqueous alteration (\citealt{sunshine-2008}, \citealt{devogele-2018}). Barbarians might represent an old population of asteroids still preserving some of the properties of the early protoplanetary disk. In addition, \cite{mahlke-2023} suggested that the variety of CV chondrites could at least partially match the spectra of Barbarians (and L- type in general), without any over-abundance of CAIs, which adds further uncertainty to the interpretation of this class of objects.

(2085)~Henan was also shown to be a good Barbarian candidate (\citealt{devogele-2018}) along with other asteroids in the region. Potentially, this Henan family could be the result of most ancient disruption of an L-type parent body known so far, so it is particularly interesting to attempt a more detailed analysis. 

To this goal we exploit the reflectance spectra released with the \gaia DR3 (\citealt{gaiacollaboration-2023}), which contains information on more than 60,000 objects. \gaia spectra at high $SNR$ have a straightforward interpretation, while the behavior of faint spectra with low $SNR$ is more complex, especially at the longest wavelengths. This limitation, which was already discussed in \cite{gaiacollaboration-2023}, will probably be overcome in future data releases.\\
The ancient L- type family identified by this work was confirmed by additional data found in the literature (taxonomy, polarimetry, spin obliquity) and by the independent V-shape detection method (\citealt{bolin-2017}, \citealt{delbo-2017}). 

The article is organized as follows. In Sect. \ref{Identification}, we describe how the L- type family has been identified using \gaia spectra, we give a first approximation of its age and we apply the V-shape identification method developed by \cite{bolin-2017} and \cite{delbo-2019} to prove that the observed V-shape is not the result of a statistical artifact. In Sect. \ref{numerical_simulation}, the results of numerical integrations trying to reproduce the observed V-shape are illustrated. In Sect. \ref{Additional Data} we show how additional data, the spin obliquity in particular, help to confirm the detected V-shape. In Sect. \ref{literature comparison} we compare the family membership obtained from \gaia spectra to HCM family memberships and to known L- types previously classified in the literature. Finally, in Sect. \ref{Conclusion}, our results and prospects are presented.

\section{Identification of the L- type family}\label{Identification}

\subsection{Identification of the family members} \label{Identification from Spectra}

\begin{figure}
   \centering
   \includegraphics[width=\hsize]{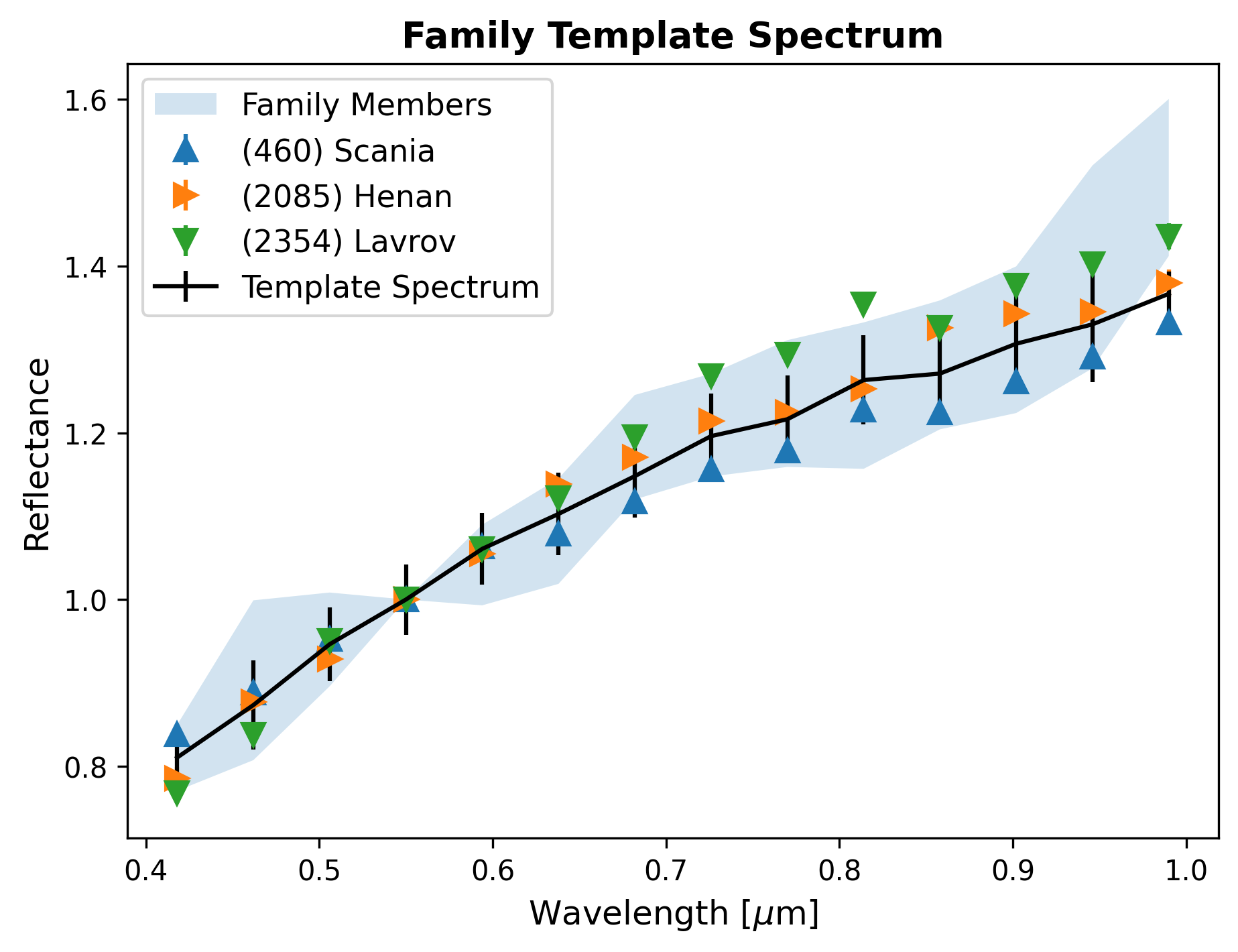}
      \caption{\small Template spectrum of the L- type family (in black). The reflectances have been computed by averaging the \gaia spectra of (460) Scania, in blue, (2085) Henan, in orange, and (2354) Lavrov, in green. The area in light blue represents the region where the spectra of the family members identified in our analysis fall within one standard deviation.}
         \label{templatespectrum}
   \end{figure}

We follow the same procedure developed in \cite{balossi-2024}, who demonstrated that asteroid families can be identified using only \gaia spectral data, successfully recovering Tirela/Klumpea and Watsonia, two well-characterized L-type families. 

We start by considering all asteroids located between semi-major axes $a_{min} = 2.50$ AU and $a_{max} = 2.85$ AU, eccentricities $e_{min} = 0.02$ and $e_{max} = 0.1$, and inclinations $\sin(i_{min}) = 0.02$ and $\sin(i_{max}) = 0.1$. The proper elements $a$, $e$ and $i$ and the absolute magnitudes $H$ were retrieved from the Asteroid Family Portal (AFP, \citealt{novakovic-2022}), while the albedos $\rho_V$ and the diameters $D$ were taken, when available, from the WISE/NEOWISE database \citep{masiero-2011}. A total of 1575 objects have been observed by \gaia within this region, 1116 of which had at least one albedo measurement reported in WISE/NEOWISE. 

\begin{figure*}[ht]
   \centering
   \includegraphics[width=\textwidth]{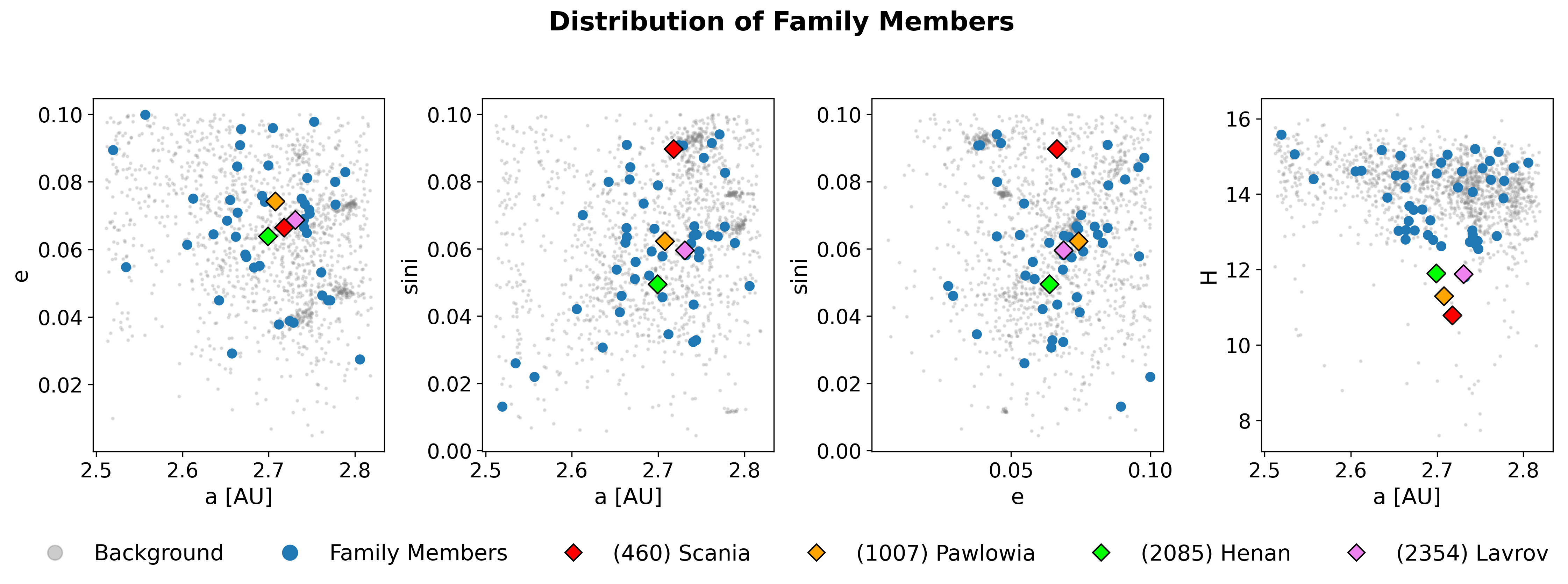}
      \caption{\small Distribution in the phase space of the proper elements of the L- types (blue circles) and all the other objects observed by \gaia (grey circles). The four largest L- types are marked by diamonds. The panels are, from left to right, ($a$, $e$), ($a$, $\sin(i)$), ($e$, $\sin(i)$) and ($a$, $H$).
      }
         \label{properlements}
   \end{figure*}

Three large confirmed L- types have reflectance spectra in \gaia DR3 within this region, namely (460) Scania, (2085) Henan, and (2354) Lavrov (these last two objects also have visible-NIR spectra in \citealt{devogele-2018}, the first has a visible-NIR spectrum in \citealt{demeo-2009}). The \gaia spectra of these three objects, which have a larger signal-to-noise with respect to the other candidate family members in the surroundings, were averaged to create a template spectrum to represent a possible L- type family. The template spectrum is shown in black in Figure \ref{templatespectrum}, where the errors represent the squared sum of the uncertainties of the spectra of all objects observed by \gaia within this region.\\
The objects were then classified into eight spectral classes according to the color taxonomy described in \cite{balossi-2024}.

A total of 55 objects were classified as L- types and were later analyzed by a $\chi^2$ similarity method, whose aim is to quantify the difference between the spectrum of a specific object and the template family spectrum. The goal is to reject possible L-types that have been wrongly assigned to the class by the classifier. We used the same definition of the $\chi^2$ as reported in \cite{balossi-2024}, with the same limit at $\chi^2=2$. Four L- types with $\chi^2>2$ were thus removed from the family. These are faint objects with low $SNR$, showing large irregular fluctuations in their spectra upon closer inspection.
   
The distribution in the phase space of the proper elements of the remaining 51 L- types is reported in Figure \ref{properlements}. The L- types are very spread in $a$, $e$, and $\sin(i)$ without any particular clustering. However, the distribution in the ($a$, $H$) plane, especially for $a > 2.6$ AU, resembles a V-shape with (460) Scania at its vertex. This seems to suggest that the 51 objects might be a family that resulted from the breakup of a large L- type in this region. Given the spread in the proper elements, this family must be very old. 

It is essential to note that some of these 51 objects may not be real family members but rather misclassified bodies or L-type interlopers. For instance, the three objects located below 2.60 AU in Figure \ref{properlements} are likely interlopers, as their positions deviate significantly from the main distribution of the family. This is particularly evident when examining their locations in the ($a$, $\sin(i)$) and ($a$, $H$) panels. \newtext{In this regard, we repeated the whole analysis using the catalog of proper elements computed by \cite{nesvorny-2024-properelements}. The membership based on this dataset is almost the same as the one determined from AFP, but it excludes from the family three small objects, two of which are among the ones below 2.60 AU. The memberships obtained from the two proper elements datasets are reported in Table \ref{tab:tablemembership}.}

We also remark here that in previous classifications, (2085)~Henan was supposed to be the largest member of the family, while here it seems that (460) Scania, not included in previous clustering results obtained by the HCM, must be considered.

\begin{figure}
   \centering
   \includegraphics[width=\hsize]{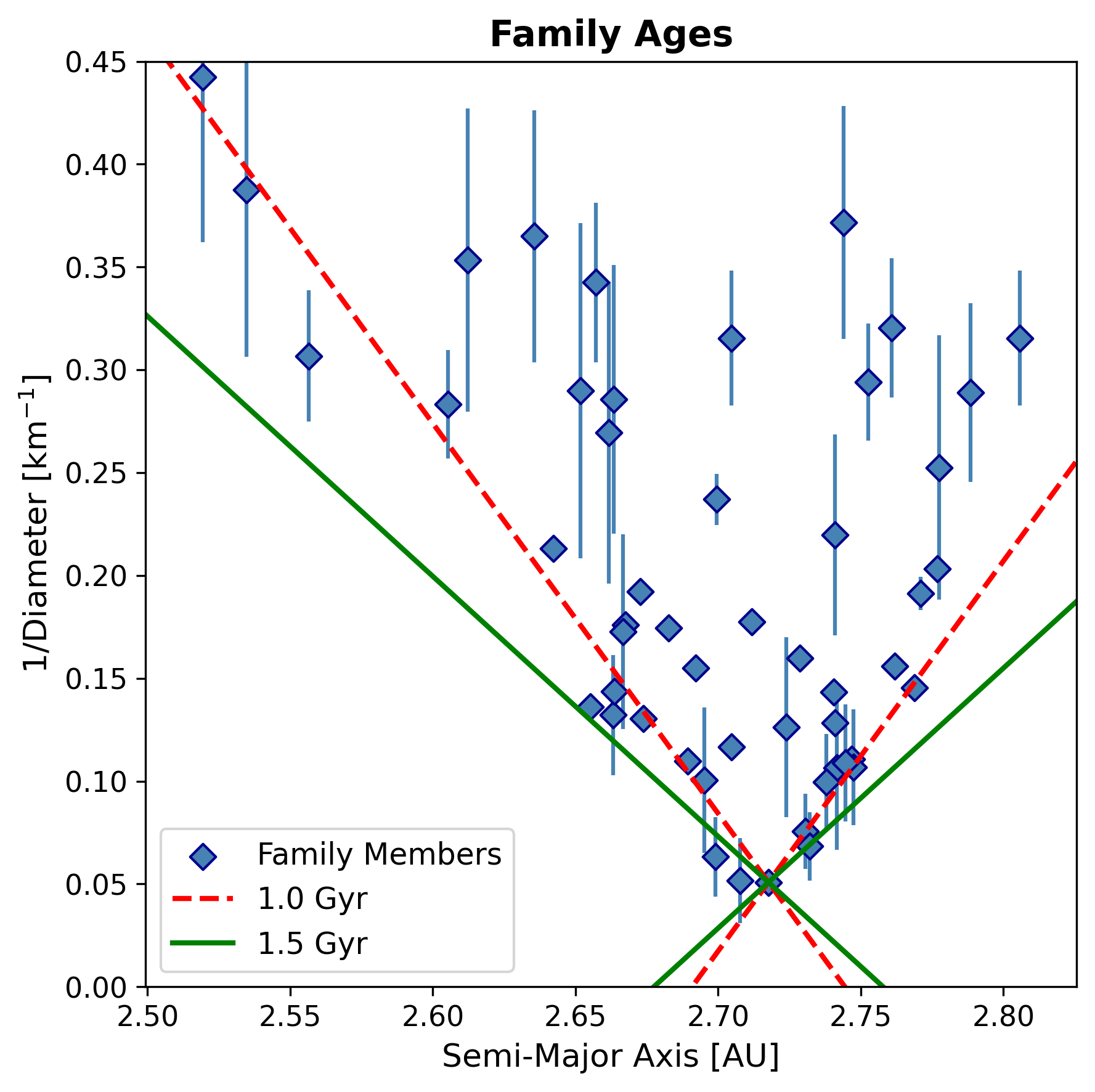}
      \caption{\small Family members (in blue) in the ($a$, $1/D$) plane fitted by V-shapes corresponding to different ages. The green continuous line corresponds to 1.5 Gyr and the red dashed line to 1.0 Gyr.}
         \label{familyages}
   \end{figure}

\subsection{V-shape detection and age determination} \label{V Shape Detection}

To obtain a first estimate of the family age we fitted the slopes of the V-shape in the ($a$, $1/D$) plane following the procedure described in \cite{spoto-2015}. Due to the limited number of objects in the family, we adopted the simplified procedure described in \cite{balossi-2024} for the Watsonia family and overplotted tentative V-shape boundaries corresponding to specific ages.

The diameters were taken from WISE/NEOWISE when available, otherwise, they were computed from $D = 1329 \cdot 10^{-H/5\sqrt{\rho_V}}$, where $H$ is the absolute magnitude of the object and $\rho_V$ the mean albedo of the family ($\rho_V =$ 0.20 $\pm$ 0.07). The distribution of the family members in the ($a$, $1/D$) plane is reported in Figure \ref{familyages}.

To determine the slopes of the V-shape corresponding to a specific age, the Yarkovsky calibration for the family must be determined, which corresponds to the value of the drift in semi-major axis $da/dt$ for a family member with unit diameter. When both $da/dt$ and the age are known, the slope of the V-shape is obtained from $S = 1 / (age \cdot da/dt)$. Unfortunately, there are no objects in this family or L- types in general with a known Yarkovsky measurement, therefore as in \cite{balossi-2024}, we decided to use the S- type (99942) Apophis as reference asteroid, whose Yarkovsky drift has been accurately computed by \cite{fenucci-2024}. We thus converted the Yarkosvky drift computed for (99942) Apophis for the L- type family using the scaling formula:

\begin{equation}
\frac{da}{dt} = \left( \frac{da}{dt} \right)_A \ \frac{\sqrt{a}_A \ (1-e_A^2)}{\sqrt{a} \ (1-e^2)} \ \frac{D_A}{D} \ \frac{\rho_A}{\rho} \ \frac{\cos(\phi)}{\cos(\phi_A)} \ \frac{1-A}{1-A_A}
\label{EqYarkCalibration}
\end{equation}

\newtext{Here $a$, $e$, $\rho$ and $A$ are respectively the semi-major axis, the eccentricity, the density, and the Bond albedo representative for the family}. When a subscript $A$ is present, the quantities are referred to (99942) Apophis (and taken from \citealt{moskovitz-2022}, \citealt{brozovic-2018}, and \citealt{berthier-2023}). 

The diameter $D=1$ km is the diameter of a fictitious unit asteroid, while $\cos(\phi)$ is the spin axis obliquity equal to $\pm 1$ depending on which side of the V-shape is considered. \newtext{For the density of (99942) Apophis, we used $\rho_A = $ 2.25 g/cm$^3$ \citep{farnocchia-2024}. The L- type density $\rho$ is instead unknown, and we considered the reference value taken from \cite{carry-2012}, which, however, is quite inaccurate due to the limited number of large L- types with accurate measurements of both mass and diameter.} We thus used the reference density corresponding to C- types, $\rho = 1.41 \pm 0.69$ g/cm$^3$, as suggested by the link between Barbarian asteroids and CO and CV meteorites, in turn related to C- types, as reported in \citet{sunshine-2008} and \citet{mahlke-2023}. 

Assuming a relative uncertainty of 0.3, we obtain a Yarkovsky calibration for the L- type family of $da/dt = (5.27 \ \pm \ 1.58) \cdot 10^{-4}$ AU/Myr.

The V-shapes corresponding to ages of 1.0 Gyr and 1.5 Gyr are reported in Figure \ref{familyages}. Given the uncertainties on the Yarkovsky calibration and on the diameters of the family members, it is difficult to determine the exact age of the family. However, it is evident that it is \newtext{quite} old, around 1.0 Gyr, which is in good agreement with the spread of the proper elements. In addition, the inner side is less inclined than the outer one, thus suggesting that the latter one might be younger. This might be explained by an anisotropy in the ejection velocity field following the breakup event, which will be better analyzed in Section \ref{numerical_simulation}. Objects on the right edge of the family may also be influenced by the strong 5:2 mean-motion resonance with Jupiter, located at 2.82 AU.

\begin{figure}
   \centering
   \includegraphics[width=\hsize]{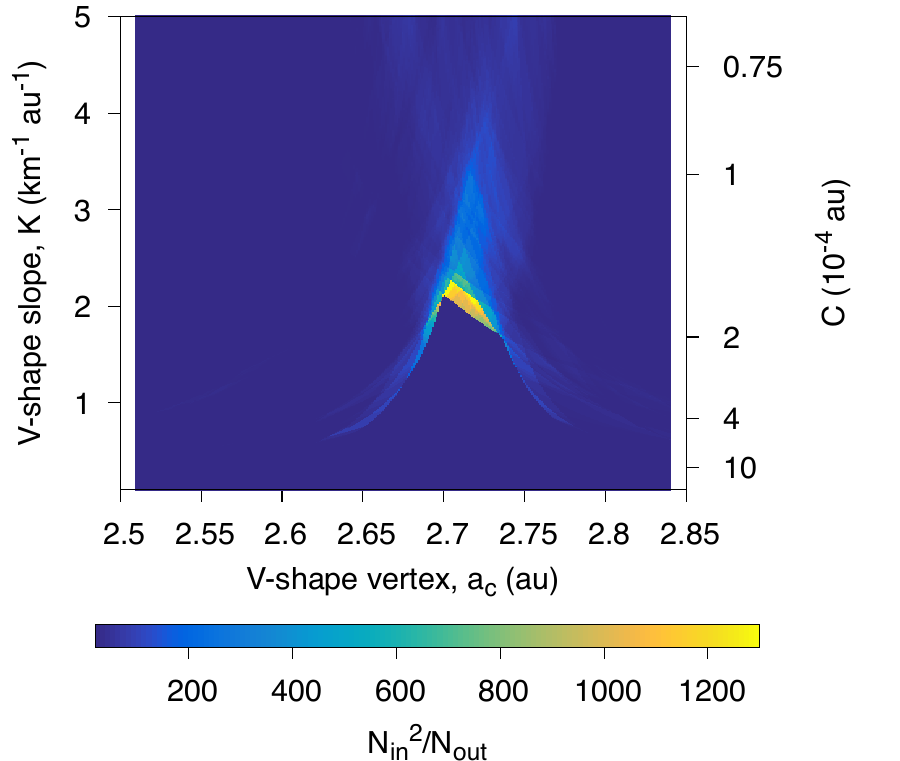}
      \caption{\small Result of the V-shape-searching method. The value of $N_{in}^2/N_{out}$ is plotted as a function of the slope ($K$) and semimajor axis of the vertex of a V-shape ($a_c$). The parameter $C$, which quantifies the width of the V shape in the ($a$, $H$) space, is computed from $C = 1/K \sqrt{p_V} / 1329$, where the geometrical visible albedo $p_V$ is set to 0.2.}
         \label{scoremap_ac_k}
   \end{figure}

To confirm the detection of the V-shape with an independent method, we apply the automated approach developed by \citet{bolin-2017}. This method searches for family V-shapes of unknown age and center ($a_c$) in a population of asteroids as a function of $a_c$ and the slope $K$ of the sides of the V. Slightly different implementations of the method exist (\citealt{bolin-2017}, \citealt{delbo-2017}, \citealt{ferrone-2023}, \citealt{walsh-2013}). Here, we use the so-called $a_w$ variant (\citealt{delbo-2017}, \citealt{ferrone-2023}): in the ($a$, $1/D$) space a nominal-V of equation $1/D = K|a-a_c|$ (where $||$ indicates the absolute value) is drawn. Next, we consider an “interior lobe” between the line $y=K(|a - a_c| + a_w)$ and the nominal line, and an “exterior lobe” between $y=K(|a - a_c| - a_w)$ and the nominal line. We count the number of bodies within the interior $N_i$ and exterior lobes $N_e$ as a function of $a_c$ and $K$. We use the quantity $N_i^2/N_e$ as a score for the detection of the V-shape. A grid of test values is created that spans a range of both $a_c$ and $K$. A score is calculated at each coordinate. Local maxima of this score map are candidate detections of family V-shapes (Figure \ref{scoremap_ac_k}).

We apply the V-shape search with $a_w=0.03$ AU on the population of L- types with 2.5 < $a_c$ < 2.8 AU and 0.2 < $K$ < 3 AU$^{-1}$km$^{-1}$. The uncertainties on $a_c$ and $K$ are estimated by a Monte Carlo approach. We performed 104 iterations where at each one we carried out a new V-shape search randomly varying the value of $a_w$ uniformly between 0.02 and 0.05 AU. Smaller values of $a_w$ produce visibly noisy $N_i^2/N_e$ score maps, while larger values of $a_w$ create inner and outer V-shape lobes that are too big compared to the area of the ($a$, $1/D$) plane to scan. At each iteration, we also vary the value of the asteroid diameters and proper semi-major axis. To do so, we added to the nominal value of the proper semi-major axis a random value normally distributed with a zero mean and a standard deviation equal to the semi-major axis uncertainty. Likewise, we added to the nominal value of the diameter a random value normally distributed with a zero mean and a standard deviation equal to the diameter uncertainty. We find that the distributions of $a_c$ and $K$ values have mean values of 2.71 AU and 2.15 AU$^{-1}$km$^{-1}$, respectively as well as standard deviations of 0.02 AU and 0.35 AU$^{-1}$km$^{-1}$, respectively. We interpret these results as the detection of a V-shape associated with an L- type family in the central Main Belt. 

We use the distribution of $a_c$ and $K$ values to estimate the age of the family from (Eq. 2 of \citealt{ferrone-2023}) also during the Monte Carlo simulation. Assuming family members' bulk densities of $1.41 \pm 0.69$ g/cm$^3$, we obtain a family age of 1.1 $\pm$ 0.6 Gyr.

\begin{figure*}[ht]
   \centering
   \includegraphics[width=\textwidth]{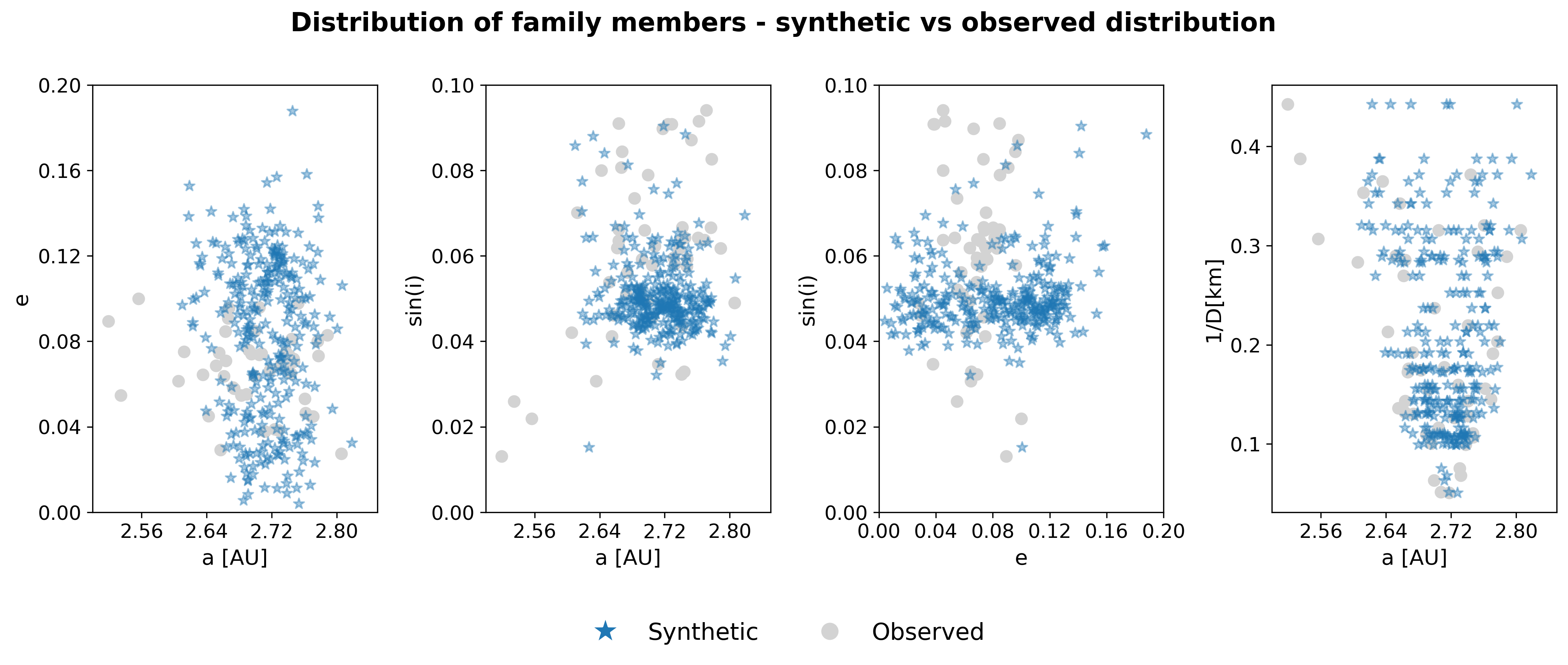}
      \caption{\small Comparison between the proper elements of the family members observed by \gaia (grey circles) and the mean proper elements of the synthetic family members integrated in \textit{REBOUND} (blue stars). The panels are, respectively, from left to right, ($a$, $e$), ($a$, $\sin(i)$), ($e$, $\sin(i)$), and ($a$, $1/D$). }
         \label{reboundintegration}
   \end{figure*}
   
\section{Numerical simulation of the family} \label{numerical_simulation}

We consider here a numerical simulation of the evolution of the family, with the goal of confirming its dispersion in the space of the osculating elements, and its age. To integrate the system, we used the N-body integrator \textit{REBOUND} \citep{rein&liu-2012} and the \textit{REBOUNDx} package to incorporate the Yarkovsky effect (\citealt{tamayo-2020}, \citealt{ferich-2022}).\

\newtext{
The semi-major axis $a_{PB}$, eccentricity $e_{PB}$, and inclination $i_{PB}$ of the parent body at the moment of breakup were taken as averages of the proper elements of the family members retrieved from \gaia. The true anomaly $f_{PB}$, the longitude of the perihelion $\omega_{PB}$ and the longitude of the ascending node $\Omega_{PB}$ were assumed to be 10, 180, and 130 degrees, respectively, because by successive attempts, we verified that these were the best values reproducing the observed clusters in the proper elements phase space. Very different values produced similar results, but a shift of the final state in $i$ and/or $e$.
}

281 fragments with the same diameter $D$ of the family members observed by \gaia were placed at the breakup position. In our simulation, we prioritized fragments with smaller diameters, including only five larger fragments to account for the presence of bigger objects. The fragments were assigned an isotropic ejection velocity field, where the value of the breakup velocity $v_{ej}$ was determined from \cite{carruba-2003}, where we assumed the density of the parent body $\rho_{PB}= 1.41$ gm/c$^3$, same as the family members, and the fraction of specific energy going into the fragmentation kinetic energy $f_{KE} = 0.01$, to maintain the family as compact as possible. The radius of the parent body $R_{PB}$ was initially determined as the sum of the size of the largest and third largest family members (\citealt{tanga-1999}), which resulted in $R_{pb}=$18 km. However, generating a synthetic size distribution using the geometric model described in \cite{tanga-1999} we found that the size distribution of the 51 family members observed by \gaia was better described by a parent body with radius $R_{pb}=$ 30 km (Figure \ref{sizedistribution}), which resulted in a mass ratio between the largest fragment and the parent body of 0.04. This is a small value, but still compatible with other families observed in the Main Belt, such as Koronis and Maria (\citealt{tanga-1999}). In our simulations we therefore assumed $R_{PB} = $ 30 km.

The fragments were then assigned an isotropic ejection velocity field with the ejection velocity randomly oriented in space. The module of the ejection velocity was modeled from $v = v_{ej} (D_{PB}/D)^{\alpha_{EV}}$ (\citealt{bolin-2018}), where $\alpha_{EV}$ was assumed to be equal to one. The Yarkovsky effect was also added to the simulation \citep{ferich-2022}. The \textit{REBOUNDx} Yarkovsky model requires as input parameters the diameters, the albedos, which were taken from NEOWISE, and the densities, assumed to be $\rho = 1.41 \pm 0.69$ g/cm$^3$ (see Section \ref{Identification}).

The system was integrated in time until 630 million years \newtext{under the gravitational forces of the giant planets and the Yarkovsky effect} using WHFast, a symplectic Wisdom-Holman integrator \citep{rein&tamayo-2015, wisdom&matthew-1991}, with a timestep of 30 days. Snapshots of the simulation were taken every 10 million years. The final state of the system is reported in Figure \ref{reboundintegration}, where the mean proper elements of the synthetic fragments are compared to the \gaia family members. \newtext{Proper elements were computed following the methodology of \cite{knevzevic&milani-2003}, which involves (i) numerical integration of asteroid orbits within a realistic dynamical model for 2 Myr in the past and 10 Myr in the future; (ii) digital filtering to remove short-period perturbations; and (iii) Fourier analysis of the filtered output to eliminate the main forced terms. Starting from the initial conditions provided by the propagated simulated orbital elements over 630 Myr, we then derived the proper elements corresponding to each set of simulated orbital elements.}

The distribution of the synthetic family members reproduces the observed one. The only objects that are not reproduced are the three below 2.60 AU, which, as already discussed in Section \ref{Identification from Spectra}, are likely interlopers. The small residual differences can be explained by the possibility that the observed family is slightly older than the synthetic one (\textasciitilde 1 Gyr vs 0.63 Gyr). In addition, it was observed that the orbital elements of the breakup position affect the positions of the synthetic clusters in the proper element phase space. The synthetic distribution is also influenced by the details of the ejection velocity field, which is far more complex than the simplified model we used.

Overall, the sides of the synthetic V-shape are in good agreement with the observed one. As mentioned in Section \ref{V Shape Detection}, the V-shape is not symmetric, but the slope of the outer side is larger than the inner one. Determining the age of the family from the Yarkovsky calibration results in an age of about 500 Myr for the outer side and an age of 1000 Myr for the inner side. This quite large variation can be explained by assuming that the initial velocity field was not isotropic and that the fragments on the inner side have been ejected faster than the fragments on the outer side. The difference in the ejection velocity needed to explain the observed difference can be computed from \cite{bolin-2018} as $\Delta v_{EJ} = \frac{n}{2} |a_1 -a_2| (\frac{D}{D_{PB}})^{\alpha_{EV}}$, where $n$ is the mean motion of the parent body, $|a_1 -a_2|$ is the difference in semi-major axis of two hypothetical objects of diameters $D$ located on the sides of the V-shape, $D_{PB}$ is the diameter of the parent body and $\alpha_{EV}=1$.

It is found that a difference of around 36 m/s in the initial ejection velocity field can explain the observed difference in slope. 

\section{Additional data - polarimetry and spin obliquity} \label{Additional Data}

\begin{figure*}[ht]
   \centering
   \includegraphics[width=\textwidth]{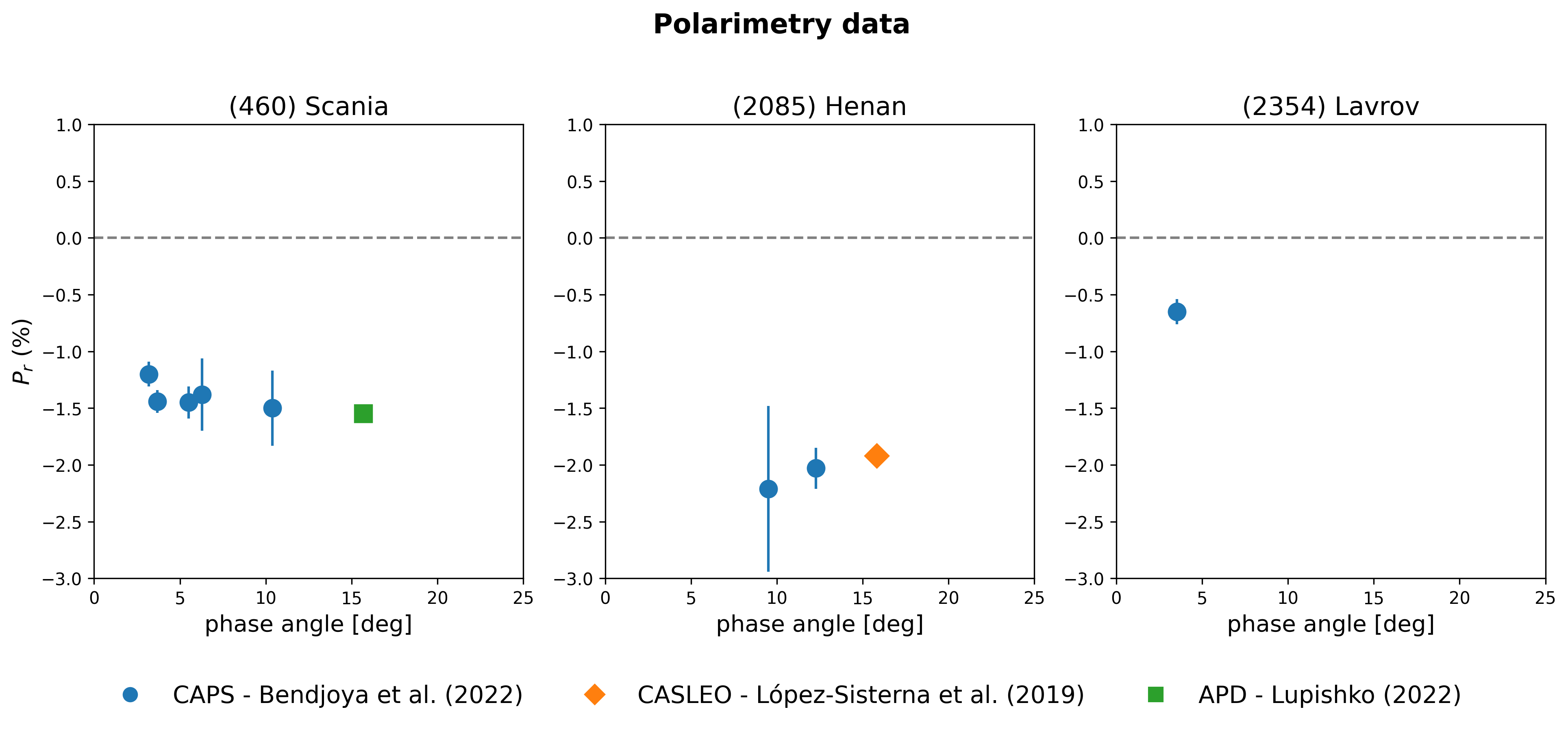}
      \caption{\small Polarimetric measurements reported in the literature for the objects belonging to the L- type family. The asteroids with data are, from left to right, (460) Scania, (2085) Henan and (2354) Lavrov.}
         \label{polarimetry}
   \end{figure*}

\begin{figure}
   \centering
   \includegraphics[width=\hsize]{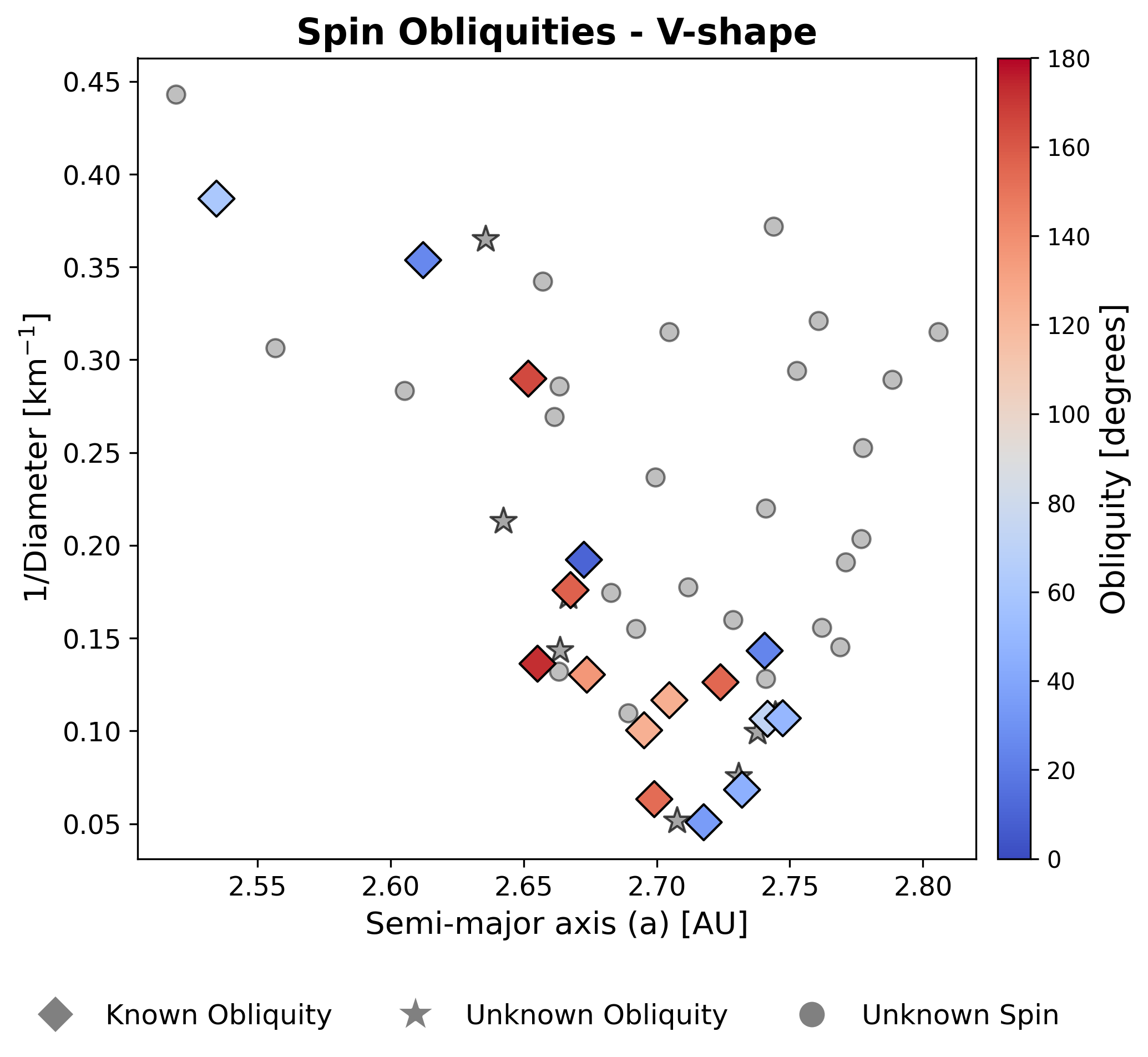}
      \caption{\small The V-shape of the family in the ($a$, $1/D$) plane with the objects marked according to the value of their spin obliquity. Grey circles are objects for which the spin is unknown. Grey stars have a known spin period, but no determination of their spin axis obliquity. Finally, diamonds have a known spin axis obliquity, which is indicated by their color: red are retrograde rotators, and blue are prograde.}
         \label{spinobliquity}
   \end{figure}

We searched the literature for additional data that could tell more about the family without being limited to the taxonomy and \gaia spectra.

Barbarian asteroids can be recognized from polarimetric measurements as they show a negative polarization at large phase angles, while normal asteroids at the same phase angles present a transition to a positive polarization (\citealt{cellino-2006}, \citealt{cellino-2014}, \citealt{gilhutton-2008}, \citealt{gilhutton-2014}).\\
Barbarian asteroids have been the subject of extensive spectroscopic studies in the literature, for example by \cite{devogele-2018}, who found that all observed Barbarians belong to the L-type spectral class. The visible and near-infrared spectra of these objects are interpreted as being consistent with a notably high abundance of spinel on their surfaces.

We searched several polarimetric databases published in the literature: the Torino catalog \citep{cellino-2005}, the Belskaya Asteroid Polarimetry \citep{belskaya-2009}, the CASLEO survey \citep{lopez-2019}, the Asteroid Polarimetric Database (APD, \citealt{lupishko-2022}), and finally the Calern Asteroid Polarisation Survey (CAPS, \citealt{bendjoya-2022}).

Only three family members have been observed: (460) Scania, (2085) Henan, and (2354) Lavrov, whose measurements are reported in Figure \ref{polarimetry}. There is just a single data point for (2354) Lavrov at a small phase angle, which is not enough to draw any conclusions about its nature. (2085) Henan has very negative polarization at large phase angles and therefore is a good Barbarian candidate, a result which was already reported in the literature, for example by \cite{devogele-2018}. However, these observations may also be consistent with the phase-polarization behavior of certain peculiar Ch- type asteroids, as illustrated in Figure 8 in \cite{bendjoya-2022}. Finally, the six measurements for (460) Scania present a large negative polarization and they cover a wide range in phase angles, strongly supporting its classification as a Barbarian.
   
We also searched the literature for the spin obliquities of the family members. The V-shape is the result of the drift in semi-major axis of an object due to the Yarkovsky effect, which depends on the orientation of its rotation axis. Prograde rotators with a spin obliquity smaller than 90 degrees drift outwards, while retrograde rotators with a spin obliquity larger than 90 degrees drift inwards. Therefore, there should be a correlation between the position of an object on the V-shape and its spin obliquity (\newtext{\citealt{vokrouhlicky-2015}}, \citealt{athanasopoulos-2022}, \citealt{athanasopoulos-2024}).

The spin obliquities were taken from three sources: the LCDB database\footnote{\url{https://sbn.psi.edu/pds/resource/lc.html}} \citep{warner-2021}, and the works by \cite{durech&hanus-2023} and \cite{cellino-2024}, who both determined the spin states using \gaia DR3 photometry. For objects without measurements in any of these works, we searched the \textit{rocks} database \citep{berthier-2023} for additional data.

We combined the data from the three studies to retrieve the spin obliquities of the L-types. For objects observed by more than one study, we adopted the following decision scheme: first, we prioritized measurements from LCDB with a good quality flag. Next, we considered the data from \cite{durech&hanus-2023}. We then included measurements from LCDB associated with a poor quality flag, and finally, we included data from \cite{cellino-2024}.

The results are shown in Figure \ref{spinobliquity}, from which it can be seen that there is a very good correlation between the obliquity and the position on the V-shape. There are only four objects that do not respect this correlation, two of which are very small and located far in semi-major axis from the rest of the family (one of these is one of the three objects below 2.60 AU already discussed in Section \ref{Identification from Spectra}). The blue point around (2.65 AU, 0.20 km$^{-1}$) might be an interloper, or its obliquity measurement might be flawed. Finally, there is a retrograde rotator around 2.72 AU, where prograde rotation is expected. However, this object lies very close to the center of the family, and its spin state may have been recently altered by collisions or other non-gravitational effects, such as the YORP effect.
   
\section{Comparison to literature} \label{literature comparison}

\subsection{Comparison to HCM algorithms} \label{Comparison to HCM}

We searched the literature to check if the L- type family had already been identified by any of the most commonly used family classification schemes, all based on HCM algorithms. In particular, we compared with Ast-DyS\footnote{\url{https://newton.spacedys.com/astdys/index.php?pc=5}} (\citealt{milani-2014}), AFP\footnote{\url{http://asteroids.matf.bg.ac.rs/fam/}} (\citealt{novakovic-2022}) and finally with the work of \newtext{\cite{nesvorny-2024}}\footnote{\url{https://sbn.psi.edu/pds/resource/nesvornyfam.html}}. Each of these classification systems reports different families in the region. 

None of the objects we classify as family members are assigned to any family by AFP, except for a few objects included in the Agnia family. These objects, however, are clear interlopers, both in spectroscopic terms, since Agnia is a S- type family, and in the phase space of the proper elements, as these objects are well separated from the core of the Agnia family. \\
Ast-DyS reports four families overlapping with our family, namely 29841, 11882, 12739, and 17392, which, however, are very small and faint clusters and have no objects in common with the L-type family.\\
\newtext{\cite{nesvorny-2024}} is the only scheme identifying an L- type family in this region, having (2085) Henan as the largest member. However, even this system fails to link to the family the larger L- types and misses the V-shape described in Section \ref{Identification}.

Instead of relying on literature classification schemes, we ran an independent HCM algorithm to identify families in this region. Even when changing the value of the limit cutoff velocity used to define families, the L- type family is never identified by the HCM. \\
This is not surprising since the family is very old and spread in the proper elements phase space, making it almost impossible for the HCM to identify it. 

Finally, \cite{mothediniz-2005} analyzed the taxonomy of asteroid families identified by their version of the HCM and they report a cluster of 37 objects with (2354) Lavrov as the largest member. Unfortunately, they do not report anywhere the full membership of the clump. However, five of these objects had available spectra at the time, plus another object linked to the clump at larger cutoff velocities. These objects are (2354) Lavrov itself, (3349) Manas, (3844) Lujiaxi, (4426) Roerich, (4726) Federer, and (5840) Raybrown, which were all classified as L- types. Three of these objects, 2354, 3844, and 5840 have also been classified as L- types by the \gaia color taxonomy and therefore included in our family. The other three objects were classified differently than the L class, even if their spectra closely resemble the family template. This evidence suggests that the color taxonomy might lose some L- types to other classes due to the intrinsic limitations of \gaia spectra and the taxonomy itself. A search in literature is thus necessary to verify how objects in this region were classified by previous works, and to understand if the family could have been identified even without \gaia spectra.

\subsection{Comparison to literature taxonomy} \label{Literature Taxonomy}

To prove that the L- type family is not the result of a bad classification of \gaia spectra, we ran an independent analysis using taxonomic classifications reported in the literature. We considered the sample observed by \gaia in this region and searched the literature for previous taxonomy classifications of these objects. For this purpose, we used the Python package \textit{rocks} (\citealt{berthier-2023}), which for each object lists all the spectral classes in which the object has been classified, along with the literature reference, the taxonomy scheme, the method of observation (photometry or spectroscopy), and the wave range of the observation (VIS, NIR or VISNIR).

For objects with multiple classifications, we considered it an L- type if more than half of the classifications corresponded to the L- class (or similar classes, such as the Ld class of \citealt{bus&binzel-2002}), giving priorities to classifications obtained from spectroscopic observations. 

\begin{figure}
   \centering
   \includegraphics[width=\hsize]{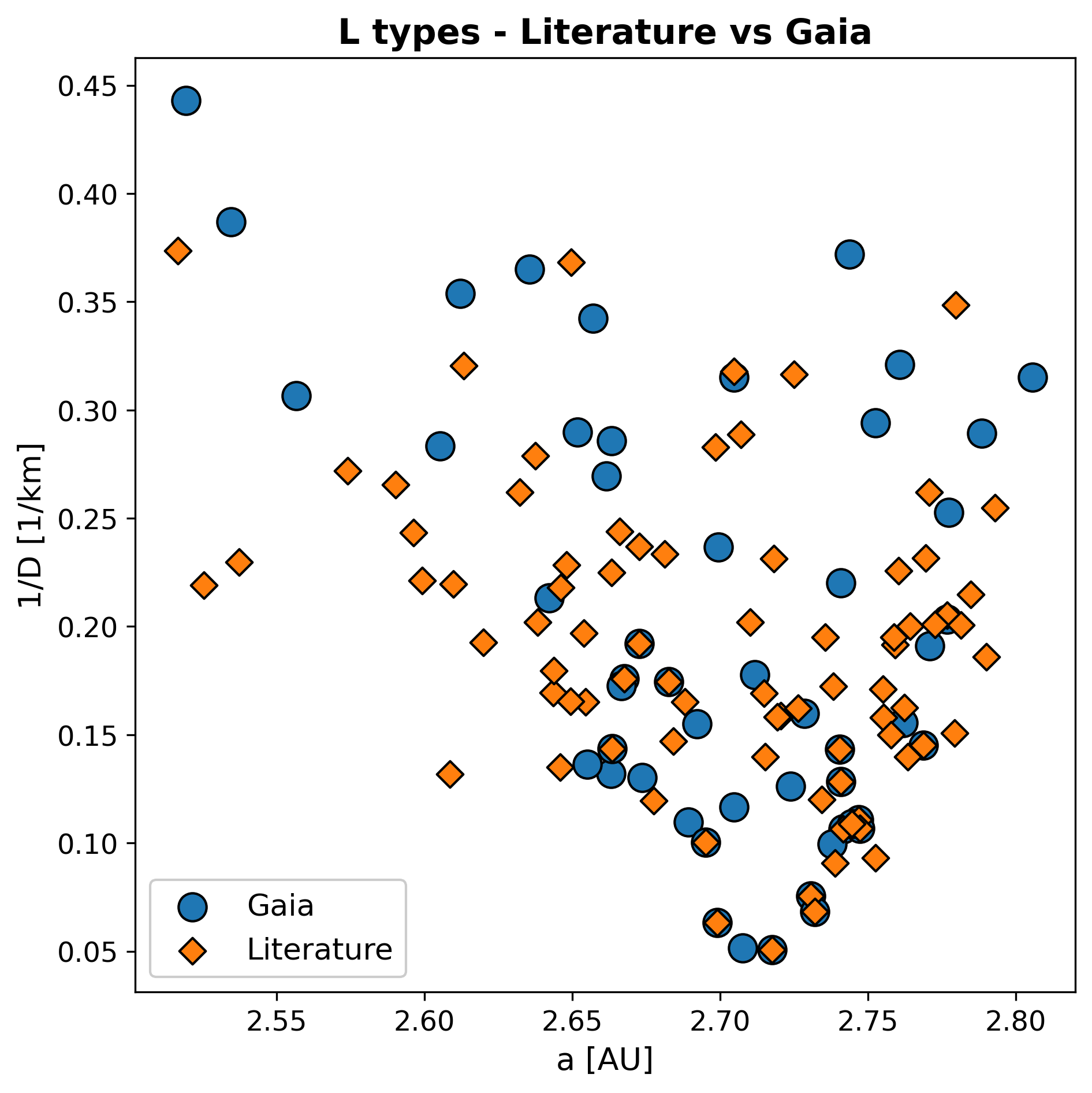}
      \caption{\small V-shape in the ($a$, $1/D$) plane of the L- types identified from the literature (orange diamonds) and from \gaia spectra (blue circles).
              }
         \label{v_shape_literature}
   \end{figure}

We found 81 objects that have been classified as L- types in the literature. These objects are shown in the ($a$, $1/D$) plane in Figure \ref{v_shape_literature}, where a V-shape is clearly recognizable. Three objects might be interlopers since they are lying outside the V-shape: the two points around (2.54 AU, 0.23 km$^{-1}$) and the point at (2.61 AU, 0.14 km$^{-1}$). An analysis of the spin obliquity showed that this last point is a prograde rotator, which is in contrast with its position on the V-shape. For the other objects there was instead a very good correspondence of the spin obliquity with the position on the V-shape, especially for the outer side. On the inner side, three small objects with diameters between 3 and 4 km turned out to be prograde rotators, thus they might be interlopers, or the measurements of their spin obliquity might be wrong.

Figure \ref{v_shape_literature} also reports the position of the L- types identified using \gaia spectra. There is a general good agreement between the L- types obtained from the literature and \gaia, especially at large diameters. The large object close to the vertex of the V-shape that is classified as L by \gaia and not by the literature is (1007) Pawlowia. This object has been classified five times in the literature, twice as S (\citealt{mahlke-2022}, \citealt{demeo&carry-2013}), twice as L (\citealt{sergeyev&carry-2021}, \citealt{carvano-2010}) and once as K (\citealt{bus&binzel-2002}). However, its albedo of 0.13 is unusually low for an S-type, thus suggesting that it is more likely associated with the M- or L- types. (1007) Pawlowia has not been included in the family because of the criteria we adopted, even if there is a good probability that it is an L- type. Additional spectroscopic observations of this object are needed to characterize its true nature. If it is indeed an L-type, the family would become particularly intriguing, as it would then feature two largest remnants with nearly identical sizes.

At smaller diameters the two V-shapes are still compatible, especially since they share the same vertex and similar slopes, although there is less agreement between the two classifications. In addition, the L- types identified from the literature seem to complete the \gaia V-shape, especially at small diameters in the inner side.

\begin{figure}
   \centering
   \includegraphics[width=\hsize]{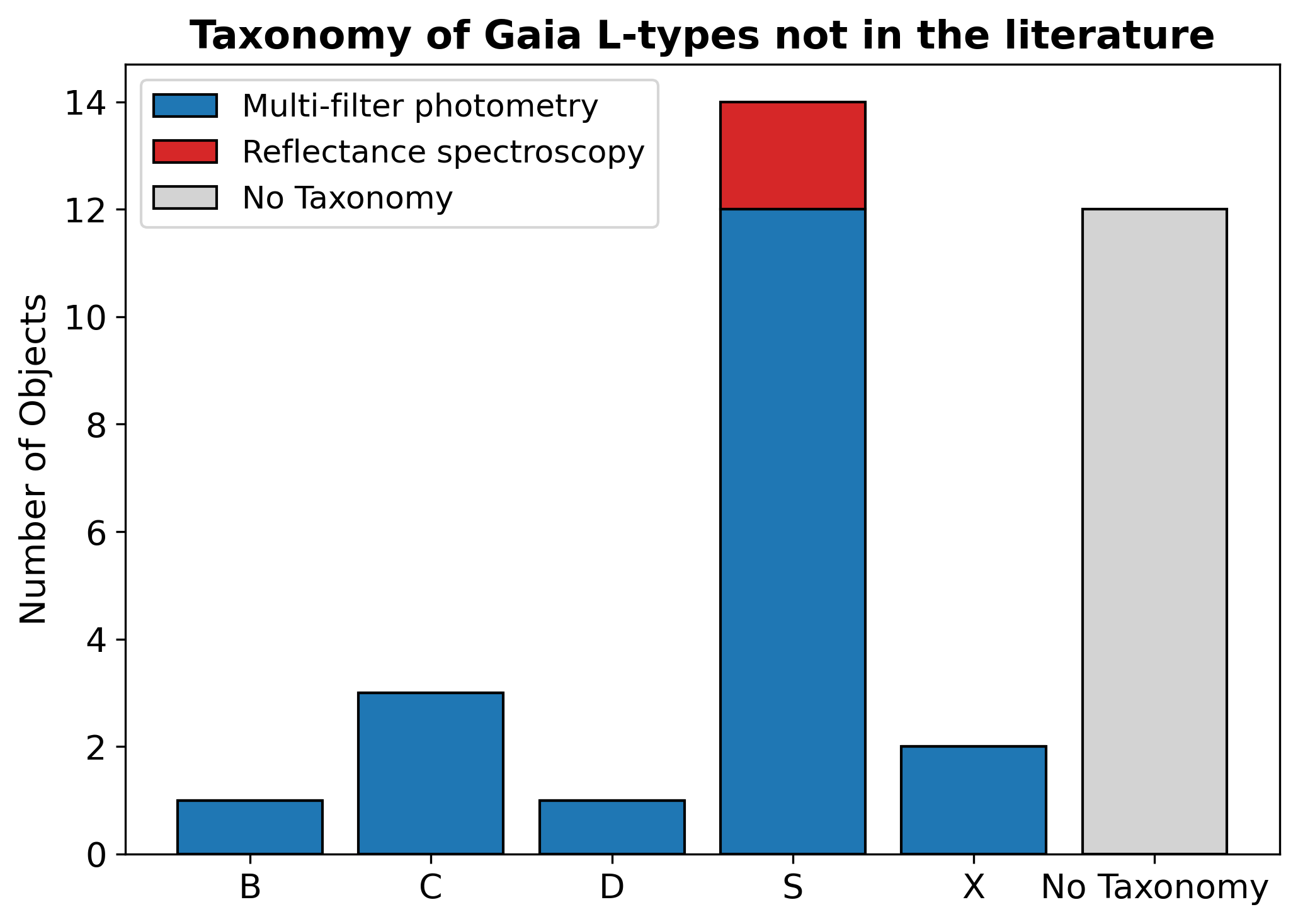}
      \caption{\small Literature taxonomy for the objects classified as L from \gaia spectra but not in the literature. In red, the objects for which the literature taxonomy has been determined from spectroscopy, in blue from photometry, and in grey objects that were never classified before this work.
              }
         \label{tax_gaia_literature}
   \end{figure}

Due to the limitations of the color taxonomy and the low $SNR$ of the \gaia spectra of small objects, it was predicted that some small objects listed as L- types in the literature would have been lost in the \gaia classification.

Less expected instead was to find objects classified as L- types from \gaia spectra but not in the literature. We therefore analyzed the literature taxonomy of these 33 objects and the results are shown in Figure \ref{tax_gaia_literature}. Two objects were classified as S- types from spectroscopic observations. One of them is (1007) Pawlowia, which was already discussed, while the other one is (4619) Polyakhova. This object was classified twice in the literature, both based on spectroscopic observations, as S- type by \cite{mahlke-2022} and as L- type by \cite{bus&binzel-2002}. (4619) Polyakhova might therefore be an L- type, but additional observations are needed to confirm its nature.\\
The classification of 19 objects was instead based on photometric observations, which are less precise than spectroscopy. Therefore, it would not be surprising if some of these objects, especially the ones classified as S- or D- types, would turn out to be L-types. The objects classified as C-, B- or X- types might be the result of a wrong classification of \gaia spectra or poor quality photometric measurements.\\
Finally, 12 objects never received a taxonomic classification before this work.

\begin{figure}
   \centering
   \includegraphics[width=\hsize]{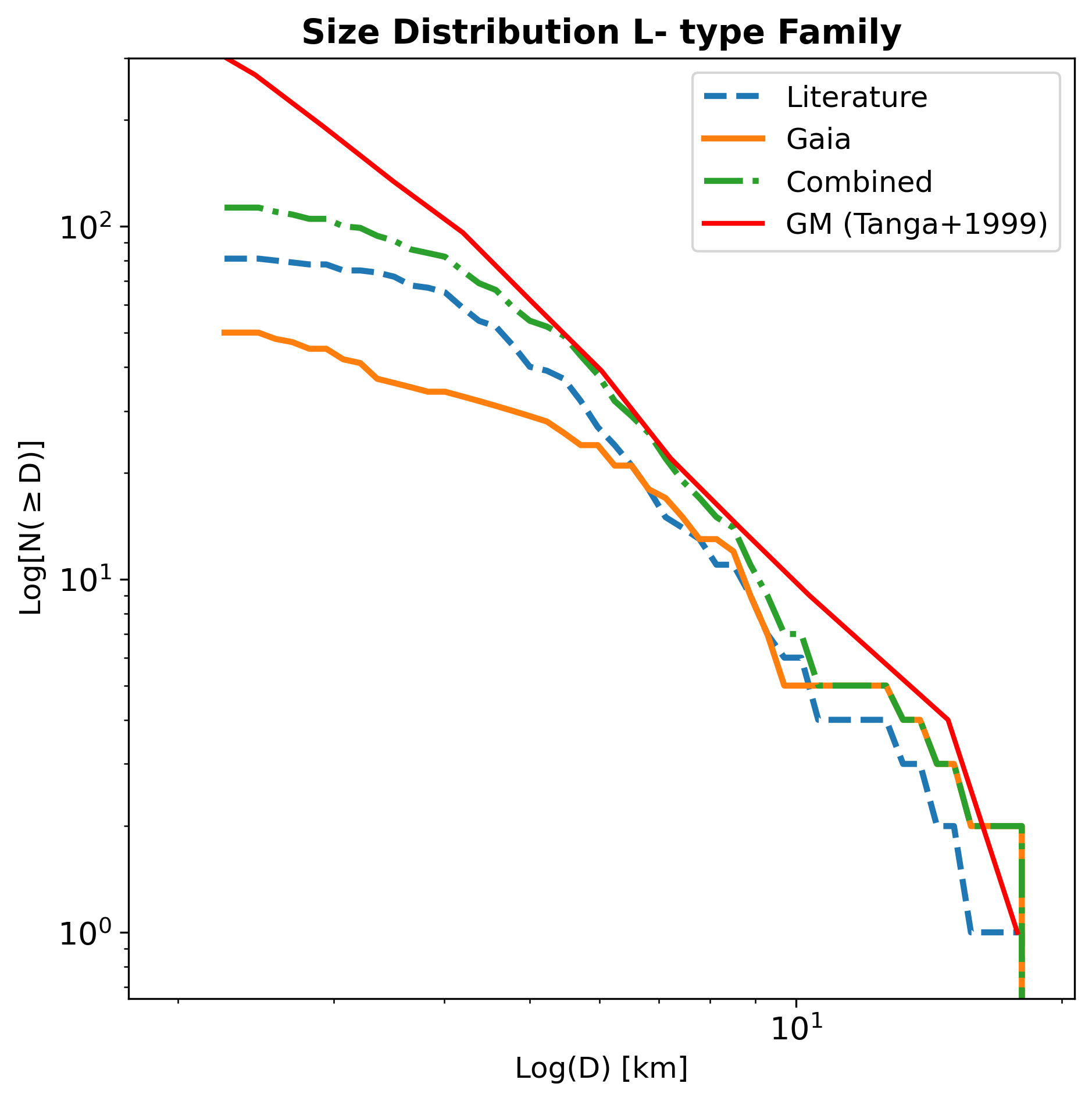}
      \caption{\small Size distributions for the \gaia membership (orange, solid line), literature membership (blue, dashed line), combined membership (green, dash-dotted line), and the geometric model by \cite{tanga-1999} (red, solid line).}
         \label{sizedistribution}
   \end{figure}

In the following, \gaia membership refers to the family membership derived from L- types identified using \gaia spectra, literature membership corresponds to the membership based on L- types reported in the literature, excluding the three interlopers discussed in Figure \ref{v_shape_literature}, and combined membership includes all L- types identified by either classification.

Figure \ref{sizedistribution} reports the size distributions of the three memberships. The \gaia membership presents an unusual size distribution due to the presence of two largest remnants (Scania and Pawlowia) of similar size. The size distribution determined from the literature is instead consistent with other known families since it does not include (1007) Pawlowia. Figure \ref{sizedistribution} also reports the size distribution obtained from the geometric model developed by \cite{tanga-1999} for a parent body of radius $R_{pb}=30$ km and a mass ratio between the largest fragment and the parent body $m_{ratio}=0.04$ (see Section \ref{numerical_simulation}). The slope of the simulated size distribution is similar to those of both \gaia and literature memberships at large diameters, while, as expected, it includes more objects at smaller diameters. Unfortunately, the geometric model does not help to clarify if (1007) Pawlowia is an interloper or an actual family member.

Overall, the agreement between the \gaia and literature memberships and the similarities between the two V-shapes show that the L- type family identified in this work is real. More observations are needed to establish the exact membership of the family at small diameters and to determine the nature of some interesting objects, (1007) Pawlowia in particular.

\section{Conclusions} \label{Conclusion}

In this paper, we have presented the identification of the family of Scania,  a L- type family in the middle Main Belt, based initially on the analysis of a taxonomy derived from \gaia DR3 spectra. Although our method solves some of the limitations of the HCM, it is not excluded that it still presents some biases, particularly since it assumes that the family is compositionally homogeneous. 

The color taxonomy that we adopt is known to have some limitations, in particular due to its tendency to classify into the most common spectral type (the S and C classes) objects that would belong to other less common spectral types. In addition, the color taxonomy might fail in classifying faint objects characterized by a low $SNR$ and flawed spectra. 

Because of these reasons, the family membership obtained from this work cannot be considered definitive. Table \ref{tab:tablemembership} in the Appendix provides a list of the L- types identified from \gaia spectra and from previous literature observations, along with their proper elements, albedos, and diameters.\\
The comparison between the \gaia and literature memberships is in good agreement, especially at large diameters. Both memberships produce a well-defined V-shape in the ($a$, $1/D$) plane, sharing the same vertex and similar slopes. The real family membership could thus be close to the combination of the \gaia and literature memberships.  We hope that with the next \gaia data release, expected for 2026, it will be possible to clarify the actual membership of this family. 

A more careful analysis is also necessary to identify any interlopers within the family. In particular, in the \gaia membership the two largest fragments present a similar size, resulting in a peculiar size distribution. The largest fragment, (460) Scania, was proven to be an L- type by previous spectroscopic observations and by polarimetry measurements. (1007) Pawlowia instead does not have any polarimetry measurement and its spectra and albedo are contradictory. Additional observations of this object are necessary to determine its true nature and to assess whether the family indeed contains two largest remnants of similar size.

An analysis of the V-shape of the family confirms that it is quite old, around 1 Gyr, and its members are very spread in the proper elements phase space. This explains why the HCM always failed to identify this family. In the best cases, small subgroups were linked together (\citealt{mothediniz-2005}, \citealt{milani-2014}).

This study confirms the interest of combining dynamical and physical properties to identify and characterize asteroid families, especially in difficult cases. Such mixed approaches are certainly going to develop to a much larger extent, with the growing set of data collected by large forthcoming surveys on fainter and fainter objects.

An open question remains the significance of the identified families composed by the rather uncommon L-type asteroids, with respect to the supposedly ancient formation age of their parent body, in an environment rich in poorly altered chondritic material and CAIs. We can speculate that, if they belong to a first generation of asteroids, only fragments of larger objects survive today, associated with dispersed and old families such as Henan and Scania. Additional high-quality spectra on smaller asteroids in the belt are needed to disentangle L-types from similar spectra and better understand this scenario.

\begin{acknowledgements}
This work presents results from the European Space Agency (ESA) space mission \gaia. \gaia data are being processed by the \gaia Data Processing and Analysis Consortium (DPAC). Funding for the DPAC is provided by national institutions, in particular, the institutions participating in the \gaia Multilateral Agreement (MLA). The \gaia mission website is https://www.cosmos.esa.int/\gaia. The \gaia archive website is https://archives.esac.esa.int/Gaia.

RB Doctoral contract is funded by Universit\`e de la C\^ote d'Azur.

This project was financed in part by the French Programme National de Planetologie, and by the BQR program of Observatoire de la C\^ote d'Azur. 

We made use of the software products: SsODNet VO service of IMCCE, Observatoire de Paris, and the associated {\tt rocks} library\footnote{\url{https://github.com/maxmahlke/rocks}} \citep{berthier-2023}; Astropy, a community-developed core Python package for Astronomy \citep{0astropy2013, 1astropy2018, 2astropy2022}; Matplotlib \citep{matplotlib_Hunter:2007}.

This work was supported by the French government through the France 2030 investment plan managed by the National Research Agency (ANR), as part of the Initiative of Excellence Université Côte d’Azur under reference number ANR-15-IDEX-01. The authors are grateful to the Université Côte d’Azur’s Center for High-Performance Computing (OPAL infrastructure) for providing resources and support.
\end{acknowledgements}

\bibliographystyle{aa}
\bibliography{references}

\appendix

\onecolumn

\section*{Appendix}

\setcounter{table}{0}

\renewcommand{\thetable}{A.\arabic{table}}

\renewcommand{\arraystretch}{1}
\begin{center}
\begin{longtable}{
  >{\centering\arraybackslash}p{2cm}
  >{\centering\arraybackslash}p{2cm}
  >{\centering\arraybackslash}p{1.5cm}
  >{\centering\arraybackslash}p{1.8cm}
  >{\centering\arraybackslash}p{1.5cm}
  >{\centering\arraybackslash}p{1.5cm}
  >{\centering\arraybackslash}p{1.8cm}
  >{\centering\arraybackslash}p{1.8cm}
} 
\captionsetup{justification=raggedright,singlelinecheck=false,labelfont=bf}
\caption{Membership for the L-type family retrieved in this work.} \label{tab:tablemembership} \\
\hline
\small \textbf{Designation} & \small \textbf{Tax. Source} & \small \textbf{H} &
\small \textbf{a [AU]} & \small \textbf{e} & \small \textbf{sin(i)} & \small \textbf{Albedo} & \small \textbf{D [km]} \\
\hline
\hline
\small 460 & \small B & \small 10.79 & \small 2.71765 & \small 0.06650 & \small 0.08977 & \small 0.26 & \small 19.69 \\
\small 1007 & \small G & \small 11.30 & \small 2.70765 & \small 0.07425 & \small 0.06228 & \small 0.13 & \small 19.37 \\
\small 2085 & \small B & \small 11.90 & \small 2.69909 & \small 0.06387 & \small 0.04945 & \small 0.19 & \small 15.81 \\
\small 2354 & \small B & \small 11.88 & \small 2.73068 & \small 0.06879 & \small 0.05957 & \small 0.19 & \small 13.23 \\
\small 3349 & \small L & \small 12.51 & \small 2.73889 & \small 0.06850 & \small 0.05779 & \small 0.15 & \small 11.01 \\
\small 3595 & \small G & \small 12.80 & \small 2.66319 & \small 0.08462 & \small 0.06621 & \small 0.26 & \small 7.57 \\
\small 3646 & \small L & \small 12.99 & \small 2.75542 & \small 0.08460 & \small 0.02556 & \small 0.15 & \small 6.33 \\
\small 3734 & \small B & \small 12.76 & \small 2.74686 & \small 0.07176 & \small 0.05753 & \small 0.20 & \small 9.04 \\
\small 3844 & \small B & \small 11.80 & \small 2.73205 & \small 0.06882 & \small 0.05809 & \small 0.17 & \small 14.64 \\
\small 4237 & \small L & \small 12.98 & \small 2.64588 & \small 0.05284 & \small 0.05032 & \small 0.21 & \small 7.41 \\
\small 4426 & \small L & \small 12.22 & \small 2.75258 & \small 0.06762 & \small 0.04675 & \small 0.21 & \small 10.75 \\
\small 4619 & \small G & \small 12.92 & \small 2.68934 & \small 0.05518 & \small 0.05214 & \small 0.16 & \small 9.12 \\
\small 4726 & \small L & \small 12.75 & \small 2.73444 & \small 0.07007 & \small 0.05746 & \small 0.25 & \small 8.34 \\
\small 4737 & \small B & \small 12.79 & \small 2.69527 & \small 0.07415 & \small 0.06604 & \small 0.14 & \small 9.96 \\
\small 4739 & \small B & \small 13.04 & \small 2.74055 & \small 0.06869 & \small 0.03233 & \small 0.25 & \small 6.98 \\
\small 4917 & \small B & \small 12.79 & \small 2.74163 & \small 0.07347 & \small 0.06672 & \small 0.17 & \small 9.39 \\
\small 5840 & \small B & \small 12.55 & \small 2.74737 & \small 0.07063 & \small 0.05932 & \small 0.21 & \small 9.36 \\
\small 6104 & \small G & \small 12.62 & \small 2.70476 & \small 0.07366 & \small 0.04565 & \small 0.24 & \small 8.58 \\
\small 6402 & \small B & \small 12.67 & \small 2.74458 & \small 0.08116 & \small 0.06421 & \small 0.20 & \small 9.19 \\
\small 6604 & \small L & \small 13.22 & \small 2.77939 & \small 0.07429 & \small 0.09925 & \small 0.21 & \small 6.63 \\
\small 7204 & \small B & \small 13.69 & \small 2.66764 & \small 0.09564 & \small 0.08437 & \small 0.18 & \small 5.68 \\
\small 7361 & \small B & \small 13.59 & \small 2.67266 & \small 0.05847 & \small 0.05105 & \small 0.28 & \small 5.20 \\
\small 7370 & \small G & \small 13.04 & \small 2.67376 & \small 0.05783 & \small 0.05615 & \small 0.19 & \small 7.67 \\
\small 7562 & \small B & \small 12.89 & \small 2.76886 & \small 0.04496 & \small 0.06372 & \small 0.28 & \small 6.88 \\
\small 7742 & \small L & \small 13.62 & \small 2.72066 & \small 0.06351 & \small 0.05519 & \small 0.20 & \small 6.30 \\
\small 7763 & \small B & \small 12.93 & \small 2.74093 & \small 0.06902 & \small 0.06393 & \small 0.22 & \small 7.80 \\
\small 7907 & \small L & \small 13.99 & \small 2.78483 & \small 0.03587 & \small 0.04794 & \small 0.21 & \small 4.65 \\
\small 8100 & \small L & \small 13.06 & \small 2.75789 & \small 0.04984 & \small 0.04012 & \small 0.25 & \small 6.67 \\
\small 8769 & \small L & \small 13.29 & \small 2.68433 & \small 0.06343 & \small 0.06270 & \small 0.15 & \small 6.80 \\
\small 8949 & \small L & \small 14.01 & \small 2.63837 & \small 0.07081 & \small 0.05089 & \small 0.20 & \small 4.95 \\
\small 9176 & \small L & \small 13.38 & \small 2.71525 & \small 0.05737 & \small 0.06291 & \small 0.18 & \small 7.15 \\
\small 9362 & \small L & \small 12.86 & \small 2.67758 & \small 0.08854 & \small 0.07445 & \small 0.21 & \small 8.36 \\
\small 9394 & \small L & \small 13.63 & \small 2.71941 & \small 0.07196 & \small 0.06478 & \small 0.18 & \small 6.32 \\
\small 9580 & \small G & \small 13.29 & \small 2.66673 & \small 0.09086 & \small 0.08070 & \small 0.25 & \small 5.79 \\
\small 9888 & \small B & \small 13.05 & \small 2.66357 & \small 0.07089 & \small 0.06358 & \small 0.25 & \small 6.97 \\
\small 9952 & \small G & \small 13.03 & \small 2.65523 & \small 0.07464 & \small 0.04121 & \small 0.23 & \small 7.34 \\
\small 10629 & \small L & \small 13.74 & \small 2.75506 & \small 0.07143 & \small 0.02279 & \small 0.14 & \small 5.85 \\
\small 10662 & \small L & \small 13.54 & \small 2.76231 & \small 0.07476 & \small 0.05691 & \small 0.15 & \small 6.16 \\
\small 10925 & \small L & \small 13.96 & \small 2.62000 & \small 0.06442 & \small 0.06426 & \small 0.22 & \small 5.19 \\
\small 11098 & \small L & \small 13.47 & \small 2.71491 & \small 0.02243 & \small 0.05393 & \small 0.21 & \small 5.91 \\
\small 11202 & \small G & \small 12.73 & \small 2.73788 & \small 0.07502 & \small 0.06165 & \small 0.16 & \small 10.05 \\
\small 11259 & \small G & \small 14.61 & \small 2.60531 & \small 0.06138 & \small 0.04209 & \small 0.20 & \small 3.53 \\
\small 12842 & \small L & \small 13.74 & \small 2.75928 & \small 0.05820 & \small 0.04853 & \small 0.21 & \small 5.22 \\
\small 12959 & \small L & \small 14.58 & \small 2.63764 & \small 0.03814 & \small 0.05793 & \small 0.24 & \small 3.59 \\
\small 13257 & \small L & \small 13.60 & \small 2.68819 & \small 0.07743 & \small 0.06746 & \small 0.19 & \small 6.05 \\
\small 14513 & \small L & \small 13.97 & \small 2.65390 & \small 0.03367 & \small 0.07939 & \small 0.30 & \small 5.08 \\
\small 16410 & \small L & \small 13.38 & \small 2.72643 & \small 0.06822 & \small 0.05965 & \small 0.21 & \small 6.16 \\
\small 17658 & \small G & \small 13.31 & \small 2.69217 & \small 0.07591 & \small 0.05925 & \small 0.24 & \small 6.45 \\
\small 18013 & \small B & \small 13.89 & \small 2.77675 & \small 0.08005 & \small 0.06665 & \small 0.20 & \small 4.92 \\
\small 18192 & \small G & \small 14.36 & \small 2.77743 & \small 0.07328 & \small 0.08262 & \small 0.20 & \small 3.96 \\
\small 18660 & \small L & \small 13.78 & \small 2.73567 & \small 0.05516 & \small 0.08098 & \small 0.21 & \small 5.13 \\
\small 18676 & \small G & \small 14.62 & \small 2.61225 & \small 0.07509 & \small 0.07011 & \small 0.38 & \small 2.83 \\
\small 19498 & \small L & \small 13.59 & \small 2.79006 & \small 0.06314 & \small 0.04952 & \small 0.24 & \small 5.37 \\
\small 19669 & \small L & \small 14.10 & \small 2.64814 & \small 0.06512 & \small 0.05784 & \small 0.23 & \small 4.38 \\
\small 20267 & \small G & \small 13.91 & \small 2.64230 & \small 0.04499 & \small 0.07999 & \small 0.27 & \small 4.69 \\
\small 21810 & \small L & \small 14.31 & \small 2.59035 & \small 0.07915 & \small 0.04679 & \small 0.18 & \small 3.77 \\
\small 23814 & \small L & \small 13.82 & \small 2.78156 & \small 0.03944 & \small 0.02277 & \small 0.16 & \small 4.98 \\
\small 23839 & \small L & \small 14.24 & \small 2.71826 & \small 0.04780 & \small 0.08217 & \small 0.22 & \small 4.32 \\
\small 23860 & \small L & \small 13.84 & \small 2.69854 & \small 0.09331 & \small 0.06796 & \small 0.47 & \small 3.53 \\
\small 24591 & \small L & \small 13.59 & \small 2.64355 & \small 0.09858 & \small 0.07716 & \small 0.20 & \small 5.91 \\
\small 25147 & \small G & \small 14.71 & \small 2.78852 & \small 0.08287 & \small 0.06178 & \small 0.20 & \small 3.46 \\
\small 25848 & \small L & \small 13.77 & \small 2.76350 & \small 0.08320 & \small 0.08729 & \small 0.14 & \small 7.15 \\
\small 28826 & \small G & \small 14.18 & \small 2.72395 & \small 0.03888 & \small 0.09083 & \small 0.08 & \small 7.92 \\
\small 29101 & \small L & \small 14.83 & \small 2.72502 & \small 0.04938 & \small 0.03427 & \small 0.21 & \small 3.16 \\
\small 29481 & \small B & \small 13.60 & \small 2.68272 & \small 0.05470 & \small 0.07347 & \small 0.21 & \small 5.73 \\
\small 29710 & \small G & \small 14.49 & \small 2.65176 & \small 0.06858 & \small 0.05384 & \small 0.26 & \small 3.45 \\
\small 31607 & \small L & \small 14.17 & \small 2.68125 & \small 0.05189 & \small 0.04943 & \small 0.21 & \small 4.28 \\
\small 31623 & \small L & \small 13.42 & \small 2.65461 & \small 0.07638 & \small 0.06335 & \small 0.21 & \small 6.05 \\
\small 31688 & \small G & \small 14.06 & \small 2.74089 & \small 0.06657 & \small 0.04345 & \small 0.20 & \small 4.55 \\
\small 32625 & \small L & \small 14.36 & \small 2.76434 & \small 0.06462 & \small 0.07257 & \small 0.18 & \small 5.00 \\
\small 32973 & \small L & \small 14.50 & \small 2.57403 & \small 0.07924 & \small 0.02837 & \small 0.21 & \small 3.68 \\
\small 33193 & \small L & \small 13.86 & \small 2.64951 & \small 0.04623 & \small 0.04587 & \small 0.07 & \small 6.04 \\
\small 33257 & \small L & \small 13.78 & \small 2.75889 & \small 0.06748 & \small 0.06661 & \small 0.21 & \small 5.13 \\
\small 33676 & \small L & \small 13.51 & \small 2.73841 & \small 0.07114 & \small 0.05888 & \small 0.21 & \small 5.80 \\
\small 35502 & \small L & \small 14.20 & \small 2.59640 & \small 0.05876 & \small 0.04206 & \small 0.22 & \small 4.11 \\
\small 35583 & \small L & \small 14.07 & \small 2.76033 & \small 0.06997 & \small 0.04093 & \small 0.23 & \small 4.43 \\
\small 36832 & \small L & \small 14.36 & \small 2.79305 & \small 0.06293 & \small 0.07950 & \small 0.21 & \small 3.92 \\
\small 36903 & \small G & \small 14.55 & \small 2.69945 & \small 0.08483 & \small 0.07890 & \small 0.13 & \small 4.22 \\
\small 37003 & \small G & \small 15.20 & \small 2.74388 & \small 0.06489 & \small 0.03288 & \small 0.20 & \small 2.69 \\
\small 37372 & \small G & \small 14.50 & \small 2.66162 & \small 0.06373 & \small 0.06189 & \small 0.20 & \small 3.71 \\
\small 37959 & \small L & \small 14.85 & \small 2.59936 & \small 0.08194 & \small 0.03040 & \small 0.11 & \small 4.52 \\
\small 38373 & \small L & \small 14.10 & \small 2.77267 & \small 0.05384 & \small 0.04670 & \small 0.15 & \small 4.98 \\
\small 38418 & \small G & \small 15.06 & \small 2.53461 & \small 0.05478 & \small 0.02599 & \small 0.29 & \small 2.58 \\
\small 44025 & \small G & \small 14.38 & \small 2.76201 & \small 0.04640 & \small 0.09152 & \small 0.11 & \small 6.42 \\
\small 44267 & \small L & \small 14.10 & \small 2.76958 & \small 0.09306 & \small 0.06723 & \small 0.26 & \small 4.32 \\
\small 45133 & \small L & \small 14.52 & \small 2.66334 & \small 0.08346 & \small 0.07358 & \small 0.16 & \small 4.44 \\
\small 46872 & \small L & \small 14.32 & \small 2.66606 & \small 0.06851 & \small 0.05232 & \small 0.20 & \small 4.10 \\
\small 47550 & \small L & \small 14.19 & \small 2.77081 & \small 0.05171 & \small 0.06276 & \small 0.23 & \small 3.82 \\
\small 48488* & \small G & \small 14.40 & \small 2.55662 & \small 0.09991 & \small 0.02192 & \small 0.32 & \small 3.26 \\
\small 49287 & \small L & \small 14.16 & \small 2.64388 & \small 0.05501 & \small 0.05311 & \small 0.14 & \small 5.57 \\
\small 51864 & \small G & \small 14.61 & \small 2.72857 & \small 0.03831 & \small 0.09080 & \small 0.08 & \small 6.26 \\
\small 52628 & \small L & \small 14.42 & \small 2.63222 & \small 0.07142 & \small 0.05812 & \small 0.21 & \small 3.82 \\
\small 53175 & \small G & \small 15.05 & \small 2.71178 & \small 0.03781 & \small 0.03466 & \small 0.06 & \small 5.63 \\
\small 53267 & \small L & \small 14.02 & \small 2.64609 & \small 0.05694 & \small 0.02871 & \small 0.21 & \small 4.59 \\
\small 53906 & \small L & \small 14.80 & \small 2.67265 & \small 0.07945 & \small 0.07009 & \small 0.14 & \small 4.22 \\
\small 55171 & \small L & \small 15.16 & \small 2.64971 & \small 0.02465 & \small 0.07919 & \small 0.21 & \small 2.71 \\
\small 58711 & \small G & \small 14.18 & \small 2.66335 & \small 0.08461 & \small 0.09099 & \small 0.23 & \small 3.50 \\
\small 58763 & \small B & \small 14.84 & \small 2.70474 & \small 0.09592 & \small 0.05784 & \small 0.20 & \small 3.17 \\
\small 59331 & \small L & \small 15.04 & \small 2.77973 & \small 0.05646 & \small 0.05199 & \small 0.21 & \small 2.87 \\
\small 61135 & \small L & \small 15.19 & \small 2.51674 & \small 0.09892 & \small 0.03935 & \small 0.21 & \small 2.68 \\
\small 62181 & \small G & \small 14.84 & \small 2.80571 & \small 0.02749 & \small 0.04898 & \small 0.20 & \small 3.17 \\
\small 62215 & \small G & \small 14.88 & \small 2.76075 & \small 0.05320 & \small 0.06420 & \small 0.20 & \small 3.12 \\
\small 64715* & \small G & \small 14.69 & \small 2.75258 & \small 0.09785 & \small 0.08717 & \small 0.20 & \small 3.40 \\
\small 70628 & \small L & \small 14.20 & \small 2.71014 & \small 0.08052 & \small 0.06713 & \small 0.18 & \small 4.95 \\
\small 70778 & \small L & \small 14.45 & \small 2.61330 & \small 0.05839 & \small 0.04418 & \small 0.32 & \small 3.12 \\
\small 75360* & \small G & \small 15.58 & \small 2.51933 & \small 0.08945 & \small 0.01311 & \small 0.20 & \small 2.26 \\
\small 81251 & \small L & \small 14.52 & \small 2.60976 & \small 0.07064 & \small 0.05369 & \small 0.14 & \small 4.56 \\
\small 88565 & \small G & \small 15.02 & \small 2.65716 & \small 0.02923 & \small 0.04609 & \small 0.20 & \small 2.92 \\
\small 89286 & \small G & \small 15.12 & \small 2.77106 & \small 0.04491 & \small 0.09406 & \small 0.06 & \small 5.23 \\
\small 104010 & \small G & \small 15.17 & \small 2.63558 & \small 0.06447 & \small 0.03070 & \small 0.20 & \small 2.74 \\
\small 104246 & \small L & \small 14.56 & \small 2.70702 & \small 0.07234 & \small 0.07066 & \small 0.23 & \small 3.46 \\
\hline
\end{longtable}

\begin{tablenotes}
    \footnotesize
    \vspace{3 px}
    \item The columns report, from left to right: identifier of the asteroid, taxonomic source, absolute magnitude $H$, proper elements (semi-major axis $a$, eccentricity $e$, inclination $\sin(i)$ from AFP, \citealt{novakovic-2022}), albedo, and diameter (from NEOWISE, \citealt{masiero-2011}). The second column, taxonomic source, indicates how the asteroids have been classified as L-types: based on Gaia spectra (G), from previous literature observations (L), or by both classifications (B). \newtext{The three objects marked by a $*$ are included in the family membership based on the AFP proper elements, but excluded from the membership based on the \cite{nesvorny-2024-properelements} catalog.}
\end{tablenotes}
\end{center}

\end{document}